\documentclass[%
 reprint,
superscriptaddress,
 amsmath,amssymb,
 aps,
pra,
]{revtex4-2}

\usepackage[utf8]{inputenc}
\usepackage[T1]{fontenc}

\usepackage{graphicx}
\graphicspath{{figures/}{paper/figures/}{multi-CD/paper/figures/}}
\usepackage[caption=false]{subfig}  
\usepackage{color}
\usepackage{colortbl}
\usepackage{xcolor}
\usepackage{soul}
\usepackage{natbib}
\usepackage{amsmath}
\usepackage{mathtools}
\usepackage{bbm}
\usepackage{physics}
\usepackage[normalem]{ulem}
\usepackage[ruled,vlined]{algorithm2e}

\usepackage[caption=false]{subfig}
\usepackage[hypertexnames=false]{hyperref}

\usepackage{enumitem}  
\usepackage{cleveref}
\usepackage{braket}
\usepackage{accents} 
\newcommand{\dbtilde}[1]{\tilde{\raisebox{0pt}[0.85\height]{$\tilde{#1}$}}}

\crefname{figure}{Fig.}{Figures}
\crefname{equation}{Eq.}{Equations}
\crefname{section}{Section}{Sections}
\crefname{appendix}{Appendix}{Appendices}

\begin{document}
\title{Counterdiabatic Driving under Variational Frame Dressing}

\author{Boxi Li}
\email{boxili@outlook.com}
\affiliation{Forschungszentrum Jülich, Institute of Quantum Control (PGI-8), D-52425 Jülich, Germany}
\affiliation{Theoretical Quantum Physics Laboratory, Cluster for Pioneering Research, RIKEN, Wakoshi, Saitama, 351-0198, Japan}
\affiliation{Quantum Information Physics Theory Research Team, Quantum Computing Center, RIKEN, Wakoshi, Saitama, 351-0198, Japan}
\author{Franco Nori}
\affiliation{Theoretical Quantum Physics Laboratory, Cluster for Pioneering Research, RIKEN, Wakoshi, Saitama, 351-0198, Japan}
\affiliation{Quantum Information Physics Theory Research Team, Quantum Computing Center, RIKEN, Wakoshi, Saitama, 351-0198, Japan}
\affiliation{Physics Department, The University of Michigan, Ann Arbor, Michigan 48109-1040, USA}
\author{Felix Motzoi}
\affiliation{Forschungszentrum Jülich, Institute of Quantum Control (PGI-8), D-52425 Jülich, Germany}
\affiliation{Institute for Theoretical Physics, University of Cologne, D-50937 Cologne, Germany}

\begin{abstract}
Counterdiabatic (CD) driving accelerates adiabatic protocols by prescribing auxiliary
control fields, but often fails to map them to physically available operations. We derive a general
formalism for such a mapping.
We formulate CD driving in a variational dressed frame,
where an unconstrained auxiliary generator reshapes the effective adiabatic problem
while simultaneously forcing the applied correction to stay restricted to the native laboratory controls.
This yields a laboratory-frame commutator equation that can be solved without
constructing instantaneous eigenstates or the adiabatic and dressed-frame
unitaries.
The additional dressed-frame freedom reveals solutions that are inaccessible in the
conventional adiabatic frame.
We illustrate this mechanism in three settings:
suppression of spectator errors in a multi-qubit chain driven by a single quadrature,
an analytical acceleration of adiabatic Bell-state preparation with a fixed entangling
interaction,
and implementation of a fast adiabatic holonomic gate in a degenerate tripod manifold with correction
pulses confined to the native couplings.
Our results provide a systematic nonperturbative framework for constructing
implementable counterdiabatic protocols beyond the conventional adiabatic frame.
\end{abstract}

\maketitle

Adiabatic control protocols are widely used because they directly prescribe how to connect quantum states through instantaneous eigenstates and are intrinsically robust to many control details.
Their usefulness is limited by the long evolution times required by the adiabatic condition,
especially in noisy intermediate-scale quantum devices where decoherence and technical noise accumulate~\cite{Albash2018Adiabatica}.
Accelerating adiabatic protocols while preserving state following is therefore a central practical goal.

Counterdiabatic (CD) driving,
introduced in~\cite{Unanyan1997Laserinduced,Demirplak2003Adiabatic},
accelerates adiabatic protocols by analytically engineering auxiliary control fields that suppress diabatic transitions and enforce finite-time state following.
It has been applied to both few-level systems~\cite{Chen2010Shortcut,Guery-Odelin2019Shortcuts}
and many-body settings~\cite{DelCampo2013Shortcuts,delCampo2012Assisted}.
Related leakage-suppression methods such as Derivative Removal by Adiabatic Gate (DRAG)~\cite{Motzoi2009Simple,Motzoi2013Improving,Theis2018Counteracting,Li2025Practical}
also use adiabatic-frame reasoning,
often perturbatively when the dynamics are not exactly solvable, e.g.~expanding the algebra using Schrieffer-Wolff theory \cite{Gambetta2011Analytic}.
Variational adiabatic-gauge-potential methods can similarly recast CD driving as an operator equation that can be solved instead variationally in a chosen ansatz, also without constructing instantaneous eigenstates~\cite{Sels2017Minimizing,Kolodrubetz2017Geometry},
which is especially useful in many-body systems.

In practice, however,
the main obstacle is often not the existence of a CD term but its implementability.
The exact conventional CD term may contain operators with little or no overlap with the available laboratory controls~\cite{Takahashi2013Transitionless}.
For example,
in a two-level Landau-Zener problem, the exact CD term is proportional to \(Y\)~\cite{Chen2010Shortcut},
so a system with only \(X\) or \(Z\) control cannot implement it directly.
More broadly,
closed-form CD solutions often rely on special Lie-algebraic structure~\cite{Torrontegui2014Hamiltonian,Martinez-Garaot2014Shortcuts,Ribeiro2019Accelerated,Opatrny2014Partial},
which makes a general constructive prescription difficult.
Several strategies address this control-mismatch problem,
including digital CD ans\"atze~\cite{Hegade2021Shortcuts,Chandarana2022Digitizedcounterdiabatic},
Magnus- and perturbative expansions~\cite{Theis2018Counteracting,Theis2016Simultaneous,Ribeiro2017Systematic},
Floquet engineering~\cite{Petiziol2018Fast,Claeys2019FloquetEngineering,Schindler2024Counterdiabatic,Petiziol2024Quantum},
effective-Hamiltonian engineering and substitute-Hamiltonian constructions~\cite{Chen2016Method,Chen2021Shortcuts},
and hybrid optimization with optimal control algorithms~\cite{Cepaite2023Counterdiabatic}.

A complementary route is to change the frame in which the CD problem is posed.
Iterating time-dependent diagonalizations leads to superadiabatic
frames~\cite{Motzoi2013Improving,Theis2018Counteracting,Deschamps2008Superadiabaticity,Ibanez2012Multiple,Ibanez2012Shortcuts,Dupays2020Superadiabatic,Santos2015Superadiabatic,Theisen2017Superadiabatica},
which can expose control directions invisible in the original adiabatic frame,
although the iterative transformations eventually
diverge~\cite{Deschamps2008Superadiabaticity}.
This motivates a more direct use of frame freedom:
instead of iterating toward higher-frame CD terms and asking only afterward
whether they are implementable, one can seek a frame where
the required correction already overlaps the accessible control space.
For the \(\Lambda\) system, dressed-state constructions for implementable
shortcuts were developed in
Refs.~\cite{Motzoi2012Controlling,Baksic2016Speeding,Li2016Shortcut,Kang2017Accelerating},
and related ideas appear in perturbative DRAG treatments for multi-photon
leakage suppression~\cite{Motzoi2013Improving,Li2024Experimental,Jesus2025Analytical}.
These constructions explicitly build the relevant instantaneous eigenstates or dressed-state
frame unitaries, which hinders systematic generalization to larger systems.
Nonetheless, these examples suggest that dressed frames can provide a practical route to
implementable corrections.

Here we generalize that idea into a laboratory-basis operator framework for dressed CD.
The dressed frame generator reshapes the adiabatic problem while the applied correction remains restricted to the available controls.
The method avoids explicit construction of instantaneous eigenstates and of the adiabatic and dressed frame unitaries.
Our examples show that it can uncover implementable corrections,
even under strongly constrained control resources that are inaccessible to conventional CD.

\paragraph{Counterdiabatic driving in dressed frames}
Consider an adiabatic evolution governed by a time-dependent Hamiltonian \(H_0(t)\).
If \(H_0(t)\) varies sufficiently slowly,
a system initialized in an instantaneous eigenstate remains on the corresponding instantaneous eigenbranch.
Counterdiabatic driving adds a correction \(H_c(t)\) such that the finite-time dynamics follows the adiabatic trajectory of \(H_0(t)\).
The implemented Hamiltonian is then \(H_{\mathrm{CD}}(t)=H_0(t)+H_c(t)\).

A convenient way to derive the conventional CD condition is to introduce a time-dependent frame transformation.
Let \(V_1(t)\) be an arbitrary smooth unitary that defines a frame transformation.
In this frame,
the Hamiltonian is
\begin{equation}
\tilde{H}_1 = \tilde{H}_0 - \tilde{A}_1 \coloneq V_1^\dagger H_0 V_1 - iV_1^\dagger \partial_t V_1
\label{eq:adiabatic frame}
\end{equation}
where the tilde denotes operators in the frame defined by \(V_1\),
and time arguments are suppressed when unambiguous.
Here
\(\tilde A_1\)
is the gauge term generated by the time dependence of the frame~\footnote{In the standard adiabatic-gauge-potential notation one often writes the correction as \(\dot\lambda A_\lambda\), where \(A_\lambda\) generates changes with respect to the control parameter \(\lambda\). Here we absorb \(\dot\lambda\) into the definition of the time-dependent generator and work directly with \(A_1=iV_1^\dagger\partial_t V_1\). This convention is more convenient for the dressed-frame construction below.}.

Conventional CD is recovered by choosing \(V_1\) to diagonalize \(H_0(t)\),
in which case \(H_c=A_1\).
Its off-diagonal part is the diabatic coupling responsible for transitions between instantaneous eigenstates.
Direct construction of \(A_1\),
however,
still requires knowledge of the instantaneous eigenbasis.
An alternative formulation is the commutator equation~\cite{Motzoi2009Simple,Gambetta2011Analytic,Sels2017Minimizing,Kolodrubetz2017Geometry}
\begin{equation}
    \left[\partial_t H_0 + i\left[A_1, H_0\right], H_0\right] = 0
    \label{eq:CD equation}
\end{equation}
which can be solved directly in the lab-frame operator basis (see Supplemental Material~\cite{supplementary}).
Solving \cref{eq:CD equation} fixes the off-diagonal part of \(A_1\),
that is, the conventional CD generator up to diagonal gauge freedom, usually for a subset of the algebra \cite{Motzoi2009Simple,Gambetta2011Analytic}.
In variational CD \cite{Sels2017Minimizing,Kolodrubetz2017Geometry},
\(A_1\) is expanded in a chosen operator basis and the residual of \cref{eq:CD equation} 
is minimized.
For linear ans\"atze,
this reduces to a quadratic minimization problem.
In practice, however, this conventional CD construction is often inapplicable:
the exact or variationally optimal CD term frequently has little overlap with the
available controls~\cite{Guery-Odelin2019Shortcuts,DelCampo2013Shortcuts}.

\begin{figure}[t]
\centering
\includegraphics[width=\linewidth,draft=false]{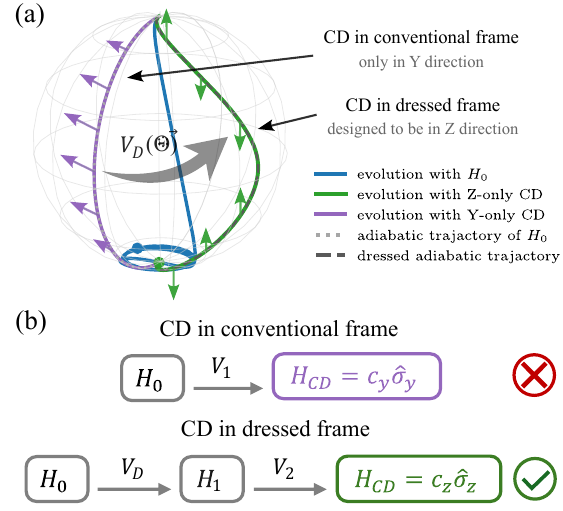}
\caption{
Overview of the dressed CD protocol.
(a)
For a Landau-Zener problem, conventional CD requires an additional \(Y\)-only correction, whereas dressed CD realizes it with an accessible \(Z\)-only control. The curves show the evolution trajectories, and the arrows indicate the control direction. The variational dressed-frame transformation \(V_D(\vec{\Theta})\) reshapes the effective Hamiltonian and the adiabatic eigenstate trajectory.
(b)
Conventional CD works directly with \(H_0\). Dressed CD instead introduces a dressed Hamiltonian \(H_1\), from which an implementable correction is constructed.
}
\label{fig:overview-dressed-frame}
\end{figure}

Here we develop a general dressed-frame approach.
Instead of requiring \(A_1\) to diagonalize \(H_0\),
we introduce a freely chosen dressed frame generator \(A_D\) (corresponding to the frame transformation \(V_D\)) that defines a new effective Hamiltonian \(\tilde H_1(t)=\tilde H_0(t)-\tilde A_D\).
In this dressed frame,
the eventual CD correction may acquire better overlap with the available controls since the frame unitary can be chosen freely.
The basic contrast with the conventional construction is illustrated in \cref{fig:overview-dressed-frame}.

We next introduce a second time-dependent unitary \(V_2(t)\),
chosen to diagonalize \(\tilde H_1(t)\).
The Hamiltonian in this dressed frame is
\(
    \dbtilde H_2
    =
    V_2^\dagger \tilde H_1 V_2
    - iV_2^\dagger\partial_t V_2 ,
    \label{eq:second-frame-tilde}
\) where the latter term is now the error.
Using \(\tilde H_1\) as the reference Hamiltonian gives the dressed-frame commutator equation
\begin{equation}
    \left[
    \partial_t\tilde H_1
    +i[\tilde H_c,\tilde H_1],
    \tilde H_1
    \right]=0.
    \label{eq:second-frame-tilde-commutator}
\end{equation}
For practical calculations,
we again want to avoid constructing dressed-frame operators explicitly.
Transforming \cref{eq:second-frame-tilde-commutator} to the laboratory operator basis gives~\cite{supplementary}
\begin{equation}
    \begin{aligned}
        [G_2,H_1] \coloneq [\partial_t H_1+i[A_D+H_c,H_1], H_1]=0
    \end{aligned}
    \label{eq:method3-G2}
\end{equation}
with \(H_1=H_0-A_D\).
Our central result is the nonlinear dressed-frame formulation, which jointly
determines the frame generator \(A_D\) and constrained \(H_c\), with the
freedom in \(A_D\) opening new directions for implementable corrections.
The special case \(G_2=0\) recovers the Lewis-Riesenfeld invariant
equation~\cite{Lewis1969Exact,Torrontegui2014Hamiltonian}.

Because \(H_1\) itself depends on \(A_D\),
the resulting equation is nonlinear.
This nonlinearity is essential: it allows \(A_D\) to deform the effective frame
so that the dressed CD correction acquires overlap with accessible controls.
The dressed frame generator \(A_D\) is not added to the laboratory Hamiltonian,
and therefore its basis operators can be chosen freely.

For complicated problems, a variational method can be built by choosing a finite set of operator bases
\begin{equation}
    A_D(t)=\sum_{\mu} a_{\mu}(t) V_{\mu},
    \qquad
    H_c(t)=\sum_{\nu} c_{\nu}(t) C_{\nu},
    \label{eq:method3-ansatz}
\end{equation}
where \(V_{\mu}\) are basis operators for the dressed frame generator and \(C_{\nu}\) are laboratory control operators.
In the variational implementation, the coefficients \(a_\mu(t)\) and
\(c_\nu(t)\) are expanded in smooth finite time bases with boundary
constraints,
so that time derivative is analytically solved and the remaining variational search is a polynomial optimization problem.
A direct measure of the residual is
\begin{equation}
    \mathcal S_{\mathrm{op}}
    =
    \int dt\,
    \norm{[G_2,H_1]}^2,
    \label{eq:method3-action-operator}
\end{equation}
which vanishes when the dressed-frame condition is satisfied.

We now turn to three applications that show how the dressed-frame construction can uncover new solutions that are inaccessible in the conventional CD approach.

\paragraph{Single-quadrature multi-spectator protection}
We consider a multi-qubit setting where a common \(X\) drive acts on one resonant
target subspace, \(\Delta_0=0\), and \(n\) spectator qubits with distinct static detunings
\(\Delta_j\).
The Hamiltonian is
\begin{equation}
    H_0(t)
    =
    \frac{\Omega(t)}{2}\sum_{j=0}^n X_j
    +
    \sum_{j=1}^n \frac{\Delta_j}{2}Z_j ,
    \label{eq:single-qubit-H0}
\end{equation}
where \(\Omega(t)\) is a smooth pulse satisfying 
\(\int_0^T \Omega(t)\,dt=\pi\). Here \(X_j\), \(Y_j\), and \(Z_j\) denote Pauli operators on qubit \(j\).
This model captures spectator-qubit crosstalk~\cite{Motzoi2013Improving,Wang2025Suppressing,Niu2024Multiqubit,Piltz2014Trappedionbased,Fang2022Crosstalk}
and tunable-coupling architectures where complex multi-quadrature pulses are
infeasible~\cite{Rimbach-Russ2022Simple,Polat2026Pulse}.

For a single spectator subspace, 
the conventional CD correction for \cref{eq:single-qubit-H0} is
\(H_{\mathrm{CD}}^{\mathrm{conv}}(t)=H_0(t)+(c_y(t)/2)Y\) with
\(c_y(t)=\Delta\dot\Omega/(\Delta^2+\Omega^2)\)~\cite{Chen2010Shortcut},
which 
requires unavailable \(Y\) control.
We instead impose \(H_c(t)=(c_x(t)/2)X\),
while allowing the dressed frame generator \(A_D\) to contain arbitrary
two-level operators.
This admits exact analytical solutions.
Parameterizing the dressed Hamiltonian as
\(
H_1(t)=\frac{\Delta}{2}[x(t)X+y(t)Y+z(t)Z]
\)
with \(x^2+y^2+z^2=1\),
the exact condition \([G_2,H_1]=0\) reduces to a single-parameter equation (see Supplemental Material~\cite{supplementary}).
For a given time-dependent profile \(x(t)\),
\begin{equation}
    y(t)=-\frac{\dot x(t)}{\Delta},
    \qquad
    z(t)=\sqrt{1-x(t)^2-\frac{\dot x(t)^2}{\Delta^2}},
    \label{eq:single-qubit-yz}
\end{equation}
the corresponding \(X\)-only correction is
\begin{equation}
    c_x(t)=\frac{\ddot x(t)+\Delta^2x(t)}{\Delta z(t)}-\Omega(t).
    \label{eq:single-qubit-u}
\end{equation}
Hence any smooth \(x(t)\) satisfying the regularity condition
\(
1-x(t)^2-\dot x(t)^2/\Delta^2>0
\)
and the endpoint conditions yields an exact \(X\)-only dressed-frame correction.

\begin{figure}[t]
\centering
\includegraphics[width=\linewidth,draft=false]{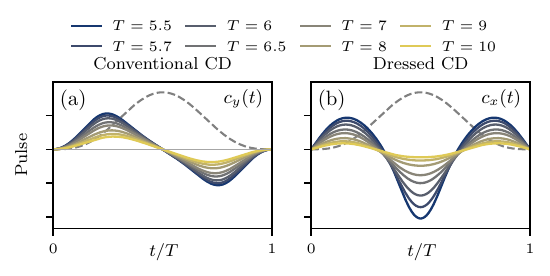}
\includegraphics[width=\linewidth,draft=false]{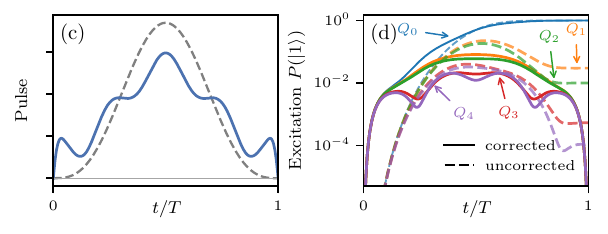}
\caption{
(a,b)~Single-qubit dressed CD for a single spectator:
(a) conventional \(Y\)-only CD coefficient \(c_y(t)\) and
(b) dressed \(X\)-only correction \(c_x(t)\) at different pulse durations
\(T\) (given in the unit of \(1/\Delta_1\)), with the gray dashed curve showing
the bare drive \(\Omega(t)\) at \(T=7\).
(c,d)~Multi-spectator generalization at \(T=10\):
(c) numerically constructed exact solution and
(d) population dynamics for four spectators
(\(\Delta_j=\{1.0,-1.1,1.9,-2.05\}\)) and one target qubit
(\(\Delta_0=0\)), where dashed lines show the bare dynamics and solid lines
the dressed-CD corrected evolution.
The target qubit undergoes a \(\pi\) pulse while all spectators remain
in \(\ket{0}\) after correction.
}
\label{fig:variational-single-qubit-time-sweep}
\end{figure}
\Cref{fig:variational-single-qubit-time-sweep}(a,b) show an example of the resulting pulse families
for a range of \(T\).
The dressed CD correction recovers an exact implementable protocol even
though the conventional CD term is orthogonal to the available control space.

This single-qubit construction extends directly to multiple spectators.
Each spectator individually satisfies \cref{eq:single-qubit-yz,eq:single-qubit-u},
but the same physical correction \(c_x(t)\) must simultaneously close all
Bloch-vector trajectories.
Inverting \cref{eq:single-qubit-u} gives a second-order ODE for each spectator,
which reduces to a coupled first-order boundary-value problem on the
unit disks driven by a single \(c_x(t)\).
The explicit ODE system and numerical details are given in
Supplemental Material~\cite{supplementary}.
\Cref{fig:variational-single-qubit-time-sweep}(c,d) show an example solution where the target qubit executes a clean \(\pi\) pulse
while all spectators remain in \(\ket{0}\).
While the conventional CD correction would require distinct correction pulses on every spectator, this dressed CD solution completely protects all spectators simultaneously.

\paragraph{Bell-state preparation by dressed CD}
We next consider Bell-state preparation in a transverse-field Ising dimer,
where dressed CD variationally finds an \(X\)-only protocol even though conventional CD contains only two-qubit operators~\cite{Takahashi2013Transitionless,Damski2014Counterdiabatic,delCampo2012Assisted}.
We consider
\begin{equation}
    H_0(t)
    =
    J Z_1 Z_2
    +
    h_x(t)X_s
    \label{eq:ising-2q-H0}
\end{equation}
with \(J=-1\) and
\(
    X_s
    =
    (X_1+X_2)/\sqrt{2}
\).
This two-qubit setting is a standard entanglement testbed for Ising interactions~\cite{Terzis2004Entanglement} and can also be viewed as the smallest building block of larger transverse-field Ising annealing architectures~\cite{King2022Coherent}.
The transverse field is
\(
    h_x(t)
    =
    l_0 + (l_f-l_0)\sin^2[\frac{\pi}{2}\sin^2(\frac{\pi t}{2T})],
\)
with \(l_0=10\) and \(l_f\) a small positive value that selects the branch approaching \(\ket{\Phi^+}\).
so it is smoothly ramped from a large positive value to a small final value.
The instantaneous ground state then evolves from a separable transverse-field state toward
the ferromagnetic Bell state \(\ket{\Phi^+}=(\ket{00}+\ket{11})/\sqrt{2}\).

\begin{figure}[t]
\centering
\includegraphics[width=\linewidth,draft=false]{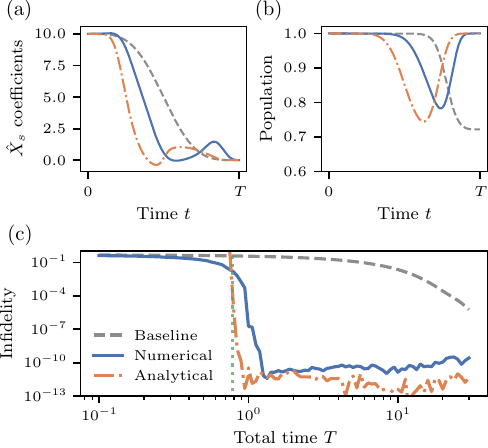}
\caption{
Accelerated Bell-state preparation using an \(X\)-only dressed CD.
In all panels, the gray dashed, blue solid, and orange dash-dotted curves denote
the baseline drive and the variational dressed-CD protocols solved numerically and analytically, respectively.
(a)
Implemented \(X\) pulse at \(T=2\).
(b)
Instantaneous ground-state population.
(c)
Final error versus total time \(T\).
The dotted vertical line marks the ideal entangling limit \(\pi/(4|J|)\).
}
\label{fig:variational-ising-2q-x-only}
\end{figure}

The conventional CD term is the two-qubit operator
\(
    ZY_s
    =
    (Z_1Y_2+Y_1Z_2)/\sqrt{2}
\).
We therefore solve \cref{eq:method3-G2} variationally with
\(
    A_D(t)=a_{ZY}(t) ZY_s+b_x(t)X_s
\)
and
\(
    H_c(t)=c_x(t)X_s
\).
As shown in \cref{fig:variational-ising-2q-x-only}, although \(Z_1Z_2\) is fixed, the variational dressed-frame protocol finds a numerically exact Bell-state preparation trajectory once sufficient entangling phase has accumulated.
Since \(Z_1Z_2\) is the sole entangling resource, a theoretical speed limit exists, \(T_{\mathrm{ent}}=\pi/(4|J|)\), marked by the green dotted line.
The finite Fourier ansatz and discrete time sweep smooth this sharp threshold into the crossover seen in the figure.
This example shows how the dressed-frame construction can exploit the existing entangling interaction while activating only the available single-qubit control.

This result reflects a hidden effective two-level structure:
within the subspace spanned by \(\ket{\Phi^+}\) and
\(\ket{\Psi^+}=(\ket{01}+\ket{10})/\sqrt{2}\),
the operators \(Z_1Z_2\), \(X_s\), and \(ZY_s\) realize an effective qubit
algebra~\cite{supplementary}, reducing the problem to a two-level
Landau-Zener form~\cite{Barnes2013Analytically,Vitanov2015Designer}
and yielding analytical pulse families benchmarked in
\cref{fig:variational-ising-2q-x-only}.
Both solving the variational equation numerically and analytically yield exact CD protocols,
indicating that the dressed-frame construction can detect and exploit such hidden structure automatically.

\paragraph{Holonomic gate in a tripod system}
\begin{figure*}[t]
\centering
\includegraphics[width=\textwidth,draft=false]{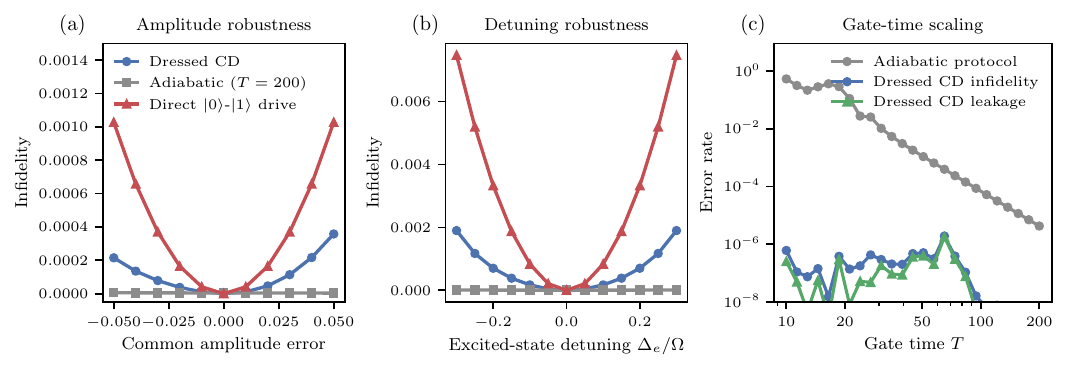}
\caption{
Tripod holonomic gate under the native-control constraint.
(a) Common rescaling of all tripod legs at fixed \(T=25\).
(b) Excited-state detuning at the same \(T\).
In both cases the dressed-frame protocol follows the robustness trend of the adiabatic holonomic loop and remains well below the external direct \(\ket{0}\)-\(\ket{1}\) benchmark.
(c) Gate-time scaling of the bare and dressed protocols.
The bare loop reaches high fidelity only in the deeply adiabatic regime,
whereas the dressed-frame correction keeps both the total error and the leakage close to the numerical accuracy.
}
\label{fig:variational-tripod}
\end{figure*}

The previous examples still involve relatively small dynamical algebras.
We finally consider a four-level tripod system, where the available controls act only between the excited state and the three lower states
and the target operation acts on a degenerate dark manifold.
Such tripod holonomies provide a prototypical setting for geometric
control~\cite{Unanyan1999Laserdriven,Leroux2018NonAbelian,Xu2020Experimental}
and generalize the STIRAP problem of
Refs.~\cite{Unanyan1997Laserinduced,Baksic2016Speeding} to a degenerate dark manifold.
In the basis \(\{\ket{0},\ket{1},\ket{a},\ket{e}\}\),
the reference Hamiltonian is
\begin{equation}
    H_0(t)
    =
    \Omega_0(t)X_{e0}
    +
    \Omega_1(t)X_{e1}
    +
    \Omega_a(t)X_{ea},
    \label{eq:tripod-H0}
\end{equation}
where \(X_{ej}=\ket{e}\bra{j}+\ket{j}\bra{e}\).
We use the smooth closed loop
\begin{equation}
    \Omega_0
    =
    \Omega \sin\theta \cos\phi,
    \quad
    \Omega_1
    =
    \Omega \sin\theta \sin\phi,
    \quad
    \Omega_a
    =
    \Omega \cos\theta,
\end{equation}
with \(\theta(t)=\theta_{\max}\sin^3(\pi t/T)\) and \(\phi(t)=2\pi t/T\).
The cubic sine envelope makes both \(\dot\theta\) and \(\ddot\theta\) vanish at the protocol endpoints,
consistent with the boundary constraints used for the dressed-frame pulses.
This loop implements a holonomic \(R_y(\pi/2)\) gate in the dark subspace~\cite{Zhang2015Fast}.
The conventional CD term for this protocol acts through the direct ground-manifold couplings,
so it lies outside the native tripod star graph.
We instead restrict the constrained physical correction to the native laboratory controls,
\begin{equation}
    H_c(t)
    =
    \Omega_{c,0}(t)X_{e0}
    +
    \Omega_{c,1}(t)X_{e1}
    +
    \Omega_{c,a}(t)X_{ea},
    \label{eq:tripod-Hc2}
\end{equation}
while the dressed frame generator \(A_D\) absorbs the non-native directions.

The dressed-frame equations are solved variationally and the result is shown in \cref{fig:variational-tripod}, with \(\Omega=1\) and \(\phi(t)=2\pi t/T\).
To calibrate the rotation angle,\(\theta_{\max}\) is first determined by matching the bare holonomy to the target \(R_y(\pi/2)\) gate and then refined by a short sweep.
Fidelity is evaluated from the full propagator projected between the initial and final dark manifolds.

Under small diabatic error, the solution preserves the geometric robustness of
the holonomic gate, as shown by evaluating the fidelity under a common rescaling
of all tripod legs and under excited-state detuning at \(T=25\).
Because of the symmetry of the holonomic dynamics,
the dressed-frame gate follows the same robustness trend with only modest extra
sensitivity relative to the adiabatic reference.
For comparison, we also show a directly driven \(\ket{0}\)-\(\ket{1}\)
transition as an external benchmark, since such a coupling is not native to
the tripod control graph.
The time-sweep panel shows the complementary advantage:
the bare loop becomes accurate only at large \(T\),
while the dressed-frame correction maintains near-perfect performance at much
shorter durations.
This example shows that the method is not limited to analytically tractable small-algebra problems,
but can also preserve the geometric robustness of a degenerate holonomic gate while keeping the correction inside the native control graph.

\paragraph{Discussion}

We have formulated dressed-frame CD as a laboratory-basis framework where a
virtual frame generator \(A_D\) reshapes the adiabatic problem while the physical
correction \(H_c\) remains within the available controls.
The examples show that this freedom reveals analytical and variational protocols
inaccessible to conventional CD, significantly increasing their potential utility.
After finite-basis parametrization, the dressed-frame equations become a
nonlinear minimization problem rather than the quadratic least-squares problem
of conventional variational CD.
However, the search remains tied to the
counterdiabatic structure through the commutator equation, preserving
qualitative features such as geometric robustness.

This formulation also suggests a route toward larger interacting systems.
The dressed-frame equations are written at the Hamiltonian level and can be
combined with structured many-body operator
ans\"atze~\cite{Sels2017Minimizing,Xie2022Variational,Takahashi2024Shortcuts}.
However, a clear distinction is that the spectrum and eigenstates depend on the dressed effective Hamiltonian
\(H_1=H_0-A_D\) and are not fixed by the original reference Hamiltonian \(H_0\).
A natural direction is to construct the dressed frame generator progressively,
for example through adaptive operator selection, Givens transformations, low-energy projection,
or Krylov/Lanczos-based approximations that track the dressed low-energy sector
without full diagonalization~\cite{Li2022Nonperturbative,Takahashi2024Shortcuts,Ohga2026Improving}.
Establishing such many-body implementations and clarifying when exact
dressed-frame solutions survive under locality and symmetry constraints are
important open questions.

\paragraph{Acknowledgments} We thank F. A. Cárdenas-López, José Jesus, Dimitrios Georgiadis and Sandeep Suresh for valuable discussions. This work acknowledges funding by the German Federal Ministry of Education and Research (BMBF) within the framework programme "Quantum technologies - from basic research to market" (Project QSolid, Grant No. 13N16149) and by Horizon Europe program via project QCFD (101080085, HORIZON-CL4-2021-DIGITAL-EMERGING02-10), project OpenSuperQPlus100 (101113946, HORIZON-CL4-2022-QUANTUM-01-SGA) and by the Cluster of Excellence Matter and Light for Quantum Computing (ML4Q2) EXC 2004/2 – 390534769.
F. N. is supported in part by the Japan Science and Technology Agency (JST) [via the CREST Quantum Frontiers program Grant No. JPMJCR24I2, the Quantum Leap Flagship Program (Q-LEAP), the Moonshot R\&D Grant Number JPMJMS256E, and the ASPIRE program (Grant Number JPMJAP2513)].

\clearpage

\onecolumngrid
\begin{center}
  \textbf{\large Supplementary material for "Counterdiabatic Driving under Variational Frame Dressing"}\\[.2cm]
\end{center}
\medskip

\setcounter{figure}{0}
\setcounter{section}{0}
\setcounter{table}{0}
\setcounter{page}{1}
\setcounter{equation}{0}
\renewcommand{\theequation}{S\arabic{equation}}
\renewcommand{\thefigure}{S\arabic{figure}}
\renewcommand{\bibnumfmt}[1]{[S#1]}

\twocolumngrid

\section{Derivation of the lab-frame dressed-frame commutator equations}
\label{app:commutator-derivation}

\subsection{First-frame equation}

Let \(V_1(t)\) be the unitary that defines the first time-dependent frame \eqref{eq:adiabatic frame}.
By definition,
\begin{equation}
    \tilde H_0 = V_1^\dagger H_0 V_1,
    \qquad
    \tilde A_1 = iV_1^\dagger \partial_t V_1 .
\end{equation}
Equivalently,
\begin{equation}
    H_0 = V_1 \tilde H_0 V_1^\dagger,
    \qquad
    A_1 = V_1 \tilde A_1 V_1^\dagger = i(\partial_t V_1)V_1^\dagger .
\end{equation}
Taking the time derivative of \(H_0\) gives
\begin{align}
    \partial_t H_0
    &=
    (\partial_t V_1)\tilde H_0 V_1^\dagger
    +
    V_1(\partial_t \tilde H_0)V_1^\dagger
    +
    V_1\tilde H_0(\partial_t V_1^\dagger) \notag \\
    &=
    -iA_1 H_0
    +
    V_1(\partial_t \tilde H_0)V_1^\dagger
    +
    iH_0A_1 .
\end{align}
Hence
\begin{equation}
    V_1(\partial_t \tilde H_0)V_1^\dagger
    =
    \partial_t H_0 + i[A_1,H_0] .
    \label{eq:app-first-frame-derivative}
\end{equation}
When \(V_1\) diagonalizes \(H_0\),
\(\tilde H_0\) is diagonal in the instantaneous eigenbasis of \(H_0\),
so \(\partial_t \tilde H_0\) is diagonal in the same basis and therefore
\begin{equation}\label{eq:diagframe}
    [\partial_t \tilde H_0,\tilde H_0]=0 .
\end{equation}
Transforming this back to the laboratory frame and applying \cref{eq:app-first-frame-derivative} and \(H_0=V_1\tilde H_0V_1^\dagger\),
we obtain
\begin{equation}
    G_1 \coloneq \partial_t H_0+i[A_1,H_0],
    \qquad [G_1,H_0]=0 ,
\end{equation}
which is \cref{eq:CD equation} in the laboratory operator basis. Here if $A_1$ corresponds to a physical control $H_c$ we can set them to cancel; however, that is not generally the case. Instead we define a new frame that does not obey \eqref{eq:diagframe} as follows.

\subsection{Second-frame dressed equation}

For the dressed-frame construction with arbitrary Hermitian $\tilde A_D$,
the first-frame Hamiltonian \eqref{eq:adiabatic frame} is
\begin{equation}
    \tilde H_1 = \tilde H_0 - \tilde A_D ,
\end{equation}
and the corresponding laboratory-frame operator is
\begin{equation}
    H_1 = H_0 - A_D .
\end{equation}
Let \(V_2(t)\) diagonalize \(\tilde H_1\).
Assuming the dressed-CD physical correction is \(\tilde H_c\),
the transitionless condition in that frame is
\begin{equation}
    \left[
    \partial_t \tilde H_1 + i[\tilde H_c,\tilde H_1],
    \tilde H_1
    \right]=0 .
    \label{eq:app-second-frame-tilde}
\end{equation}
To rewrite this equation without constructing \(V_1\) explicitly,
we again transform the time derivative back to the laboratory basis.
Using \(H_1 = V_1 \tilde H_1 V_1^\dagger\) and the same steps  as above,
we find  
\begin{equation}
    V_1(\partial_t \tilde H_1)V_1^\dagger
    =
    \partial_t H_1 + i[A_D,H_1] .
    \label{eq:app-second-frame-derivative}
\end{equation}
Moreover,
\begin{equation}
    H_c=V_1 \tilde H_c V_1^\dagger,
    \qquad
    H_1 = V_1 \tilde H_1 V_1^\dagger .
\end{equation}
Conjugating \cref{eq:app-second-frame-tilde} by \(V_1\) and using
\cref{eq:app-second-frame-derivative} yields
\begin{equation}
    \left[
    \partial_t H_1 + i[A_D + H_c, H_1],
    H_1
    \right]=0 ,
    \label{eq:app-second-frame-lab}
\end{equation}
which is the laboratory-frame dressed equation used in the main text.
Defining
\begin{equation}
    G_2
    =
    \partial_t H_1 + i[A_D + H_c, H_1],
\end{equation}
the dressed-frame condition is simply \([G_2,H_1]=0\).
The dressed adiabatic trajectory is defined by the eigenstates of \(H_1\).

\section{Variational implementation details}
\label{app:variational-details}

\subsection{Finite-basis parametrization and boundary conditions}

In the numerical implementation,
both the virtual dressed-frame generator and the constrained physical correction are expanded in finite operator bases,
\begin{equation}
    A_D(t)=\sum_\mu a_\mu(t)V_\mu,
    \qquad
    H_c(t)=\sum_\nu c_\nu(t)C_\nu .
\end{equation}
The operator set \(\{V_\mu\}\) is chosen as a basis for the dressed Hamiltonian \(H_1=H_0-A_D\),
whereas \(\{C_\nu\}\) contains only experimentally accessible control operators.
The scalar coefficient functions \(a_\mu(t)\) and \(c_\nu(t)\) are then represented in smooth finite time bases,
typically sine or Fourier series.
For protocols that require vanishing pulse amplitudes at the endpoints,
a pure sine basis is convenient because each basis function vanishes at \(t=0\) and \(t=T\).

The endpoint conditions are chosen so that the initial and final target eigenspaces coincide with those of the reference problem.
In the most direct implementation,
this is enforced by requiring
\begin{equation}
    A_D(0)=A_D(T)=0,
    \qquad
    H_c(0)=H_c(T)=0 .
\end{equation}

\subsection{Residual functional and optimization}

After parametrization,
the dressed-frame condition is enforced variationally through
\begin{equation}
    \begin{aligned}
        G_2(t)
        &=
        \partial_t H_1(t)
        + i[A_D(t)+H_c(t),H_1(t)], \\
        H_1(t) &= H_0(t)-A_D(t).
    \end{aligned}
\end{equation}
The exact dressed solution satisfies \([G_2,H_1]=0\) for all \(t\).
For the variational approach, we therefore minimize the integrated commutator residual
\begin{equation}
    \mathcal S_{\mathrm{op}}
    =
    \int_0^T dt\,
    \norm{[G_2(t),H_1(t)]}^2 ,
\end{equation}
or its discretized version on a uniform time grid.
In the numerical examples presented in this work,
the optimization is performed on such a uniform time grid over the full interval \([0,T]\).
For matrix representations,
the norm is evaluated as the Hilbert-Schmidt norm of the residual operator.

This choice is important.
In the usual conventional CD variational problem with fixed \(H_0\),
minimizing \(\operatorname{Tr} G_1^2\) is equivalent to minimizing diabatic couplings up to an additive diagonal contribution that is independent of the variational parameters.
In the dressed-frame problem,
however,
\(H_1=H_0-A_D\) depends on the variational frame generator.
As a result,
the diagonal part of \(G_2\) is itself frame dependent.
A naive loss based directly on \(\operatorname{Tr} G_2^2\) can therefore lower the cost by reshaping instantaneous energies rather than by suppressing the off-diagonal transitions that matter for adiabatic following.
The commutator residual \(\norm{[G_2,H_1]}^2\) isolates the condition relevant to transitionlessness and avoids this ambiguity.

Treating the Fourier coefficients as a parameter vector,
the minimization becomes a finite-dimensional polynomial optimization problem.
In the calculations reported here,
we use standard gradient-based optimizers such as BFGS.
Because the derivatives of Fourier functions are analytically known,
the gradient can be evaluated efficiently and accurately.
Because the dependence on \(A_D\) is nonlinear through \(H_1=H_0-A_D\),
the dressed-frame problem is no longer a quadratic variational solve,
in contrast to the conventional AGP minimization.

\subsection{Practical numerical workflow}
\label{app:numerical-workflow}

The numerical workflow used throughout this work is as follows.
\begin{enumerate}[label=(\roman*)]
    \item Choose the reference Hamiltonian \(H_0(t)\), the constrained physical control basis \(\{C_\nu\}\), and a symmetry-compatible virtual basis \(\{V_\mu\}\).
    \item Represent the coefficient functions \(a_\mu(t)\) and \(c_\nu(t)\) in smooth finite time bases and impose the desired endpoint constraints.
    \item Discretize the interval \([0,T]\) on a uniform grid, evaluate the residual operator \([G_2(t),H_1(t)]\) on that grid, and compute \(\mathcal S_{\mathrm{op}}\) numerically.
    \item Minimize the resulting nonlinear loss over the time-basis coefficients using a gradient-based optimizer.
    \item Validate the optimized protocol by checking both the residual size and the Schr\"odinger evolution under the implemented Hamiltonian \(H_0+H_c\).
\end{enumerate}
We refer to a solution as exact when the commutator residual vanishes analytically or reaches floating-point numerical precision while the propagated dynamics matches the intended adiabatic transport.

\section{X-only dressed-CD pulse construction}
\label{app:x-only-construction}

We detail the analytical and numerical construction of the \(X\)-only dressed-CD
pulses used in Sec.~III of the main text.

\subsection{Single-qubit analytical family}

For a single spectator with detuning \(\Delta\), the dressed Hamiltonian is
\(H_1(t)=\frac{\Delta}{2}[x(t)X+y(t)Y+z(t)Z]\) with \(x^2+y^2+z^2=1\).
The dressed-frame commutator \([G_2,H_1]=0\) forces
\begin{align}
    y(t)&=-\frac{\dot x(t)}{\Delta},\nonumber\\
    z(t)&=\sqrt{1-x(t)^2-\frac{\dot x(t)^2}{\Delta^2}},\\
    c_x(t)&=\frac{\ddot x(t)+\Delta^2 x(t)}{\Delta z(t)}-\Omega(t).
    \label{eq:supp-single-qubit-cx}
\end{align}
Any smooth \(x(t)\) with endpoint conditions \(x(0)=x(T)=\dot x(0)=\dot x(T)=0\)
and \(1-x^2-\dot x^2/\Delta^2>0\) produces an exact \(X\)-only dressed-CD protocol.

The bare drive is \(\Omega(t)=\Omega_0\sin^3(\pi t/T)\) with \(\Omega_0\) fixed by
\(\int_0^T\Omega(t)\,dt=\pi\).
We choose the same envelope for the dressed Bloch component,
\(x(t)=\alpha\sin^3(\pi t/T)\), and determine \(\alpha\) from the zero-area
condition \(\int_0^T c_x(t)\,dt=0\), which preserves the total \(X\)-drive area.
With \(x(t)\) fixed, \cref{eq:supp-single-qubit-cx} gives \(c_x(t)\) in closed form.

\subsection{Multi-spectator boundary-value problem}

For \(n\) spectators with detunings \(\Delta_j\), the shared correction
\(c_x(t)\) must close all Bloch-vector trajectories simultaneously.
To formulate this as a boundary-value problem, we invert
\cref{eq:supp-single-qubit-cx} to express \(\ddot x\) in terms of \(c_x\):
\(\ddot x = \Delta z\,(c_x+\Omega) - \Delta^2 x\).
Defining \(u_j(t)=x_j(t)\) and \(v_j(t)=\dot x_j(t)/\Delta_j\) on the unit disk
\(u_j^2+v_j^2<1\), the second-order dynamics reduces to the coupled first-order system
\begin{equation}
    \begin{aligned}
        \dot u_j &= \Delta_j\,v_j,\\
        \dot v_j &= \sqrt{1-u_j^2-v_j^2}\,\bigl(c_x+\Omega\bigr)-\Delta_j\,u_j,
    \end{aligned}
    \qquad j=1,\dots,n,
    \label{eq:supp-bvp}
\end{equation}
with boundary conditions \(u_j(0)=v_j(0)=u_j(T)=v_j(T)=0\).
A solution of \cref{eq:supp-bvp} guarantees that each spectator satisfies
\cref{eq:supp-single-qubit-cx}, so the dressed-frame condition is fulfilled for
all qubits by a single \(c_x(t)\).
This is a \(2n\)-state, one-control boundary-value problem.

To solve it numerically, we represent the shared control \(c_x(t)\) in a smooth
finite time basis satisfying the endpoint constraints and transcribe the
continuous boundary-value problem into a finite-dimensional nonlinear system.
This transcription may be carried out, for example, by direct shooting,
multiple shooting, or collocation on a time grid.
In our present implementation we use a cubic B-spline representation of
\(c_x(t)\) on a uniform knot grid and optimize the spline coefficients to
minimize the endpoint defect \(\sum_j[u_j(T)^2+v_j(T)^2]\), initialized from
the single-spectator analytical solution.

\section{Bell state dressed-CD constructions}

\subsection{Landau-Zener \(Z\)-only analytical family}
\label{app:lz-z-only}

For a Landau-Zener-type sweep with constant \(h_x>0\),
\begin{equation}
    H_0(t)=h_x X+h_z(t)Z,
    \label{eq:lz-z-only-H0}
 \end{equation}
and constrained physical correction
\begin{equation}
    H_c(t)=c_z(t)Z,
    \label{eq:lz-z-only-Hc}
\end{equation}
define the implemented detuning
\begin{equation}
    u(t)=h_z(t)+c_z(t).
\end{equation}
To construct exact dressed-frame solutions,
parameterize the dressed Hamiltonian as
\begin{equation}
    \begin{aligned}
        H_1(t)
        &=
        r(t)\Bigl[
        \sin\theta(t)\cos\phi(t)\,X
        +
        \sin\theta(t)\sin\phi(t)\,Y \\
        &\qquad\qquad
        +
        \cos\theta(t)\,Z
        \Bigr],
    \end{aligned}
    \label{eq:app-lz-H1}
\end{equation}
with \(r(t)>0\), and set \(A_D(t)=H_0(t)-H_1(t)\).
Using the Pauli commutation relations,
the dressed-frame condition \([G_2,H_1]=0\) is equivalent for every regular solution with \(r(t)>0\) to the angular equations
\begin{equation}
    \dot\theta=-2h_x\sin\phi,
    \qquad
    \dot\phi=2u-2h_x\cot\theta\cos\phi .
    \label{eq:app-lz-angular}
\end{equation}

This gives a direct Hamiltonian-level inverse construction.
The resulting angular reconstruction is closely related to the inverse-engineering approaches of
Barnes~\cite{Barnes2013Analytically} and of Vitanov and Shore~\cite{Vitanov2015Designer},
but here it is obtained from the general dressed-frame commutator equation rather than from direct solution of the physical two-level dynamics.
Choose any smooth angle \(\theta(t)\) satisfying
\begin{equation}
    \theta(0)=\theta_i,
    \qquad
    \theta(T)=\theta_f,
\end{equation}
\begin{equation}
    \dot\theta(0)=\dot\theta(T)=0,
    \qquad
    \ddot\theta(0)=\ddot\theta(T)=0,
\end{equation}
\begin{equation}
    |\dot\theta(t)|<2h_x
    \qquad
    \text{for all } t\in[0,T],
    \label{eq:app-lz-theta-bound}
\end{equation}
where the endpoint angles are fixed by the bare Hamiltonian,
\begin{equation}
    \cos\theta_i=\frac{h_z(0)}{\sqrt{h_x^2+h_z(0)^2}},
    \qquad
    \cos\theta_f=\frac{h_z(T)}{\sqrt{h_x^2+h_z(T)^2}} .
\end{equation}
Then define
\begin{equation}
    \phi(t)=\arcsin\!\left[-\frac{\dot\theta(t)}{2h_x}\right],
\end{equation}
choosing the solution with \(\cos\phi>0\), and reconstruct the implemented detuning from \cref{eq:app-lz-angular}:
\begin{equation}
    u(t)=\frac{\dot\phi(t)}{2}+h_x\cot\theta(t)\cos\phi(t).
    \label{eq:app-lz-u}
\end{equation}
The physical correction is then
\begin{equation}
    c_z(t)=u(t)-h_z(t).
    \label{eq:app-lz-cz}
\end{equation}
Finally, choose any smooth positive \(r(t)\) with
\begin{equation}
    r(0)=\sqrt{h_x^2+h_z(0)^2},
    \qquad
    r(T)=\sqrt{h_x^2+h_z(T)^2}.
\end{equation}
This guarantees \(A_D(0)=A_D(T)=0\).
Moreover, the endpoint derivative conditions imply \(\phi(0)=\phi(T)=0\) and \(\dot\phi(0)=\dot\phi(T)=0\), so
\begin{equation}
    u(0)=h_z(0),
    \qquad
    u(T)=h_z(T),
\end{equation}
hence \(c_z(0)=c_z(T)=0\).
Therefore every admissible choice of \(\theta(t)\) produces an exact endpoint-matched \(Z\)-only dressed-frame solution through \cref{eq:app-lz-H1,eq:app-lz-u,eq:app-lz-cz}.

For the symmetric sweep \(h_z(0)=-h_z(T)\),
the endpoint angle difference is
\begin{equation}
    \Delta\theta=\theta_i-\theta_f
    =
    2\arctan\frac{|h_z(0)|}{h_x},
\end{equation}
so \cref{eq:app-lz-theta-bound} implies the necessary bound
\begin{equation}
    T\ge \frac{\Delta\theta}{2h_x}
    =
    \frac{1}{h_x}\arctan\frac{|h_z(0)|}{h_x}
    \xrightarrow{|h_z(0)|\to\infty}
    \frac{\pi}{2h_x},
\end{equation}
which gives the corresponding speed limit for this reduced problem.
For any \(T\) above this bound one may choose a smooth monotone interpolation
\(\theta(\lambda)\), with \(\lambda=t/T\), satisfying
\(\theta(0)=\theta_i\),
\(\theta(1)=\theta_f\),
vanishing first and second derivatives at the endpoints,
and
\(\max_\lambda \left|\frac{d\theta}{d\lambda}\right|<2h_xT\).
This yields an explicit family of exact analytical solutions parametrized directly by the choice of \(\theta(\lambda)\).
For the figure shown below we use the explicit solution
\begin{equation}
    \theta(\lambda)
    =
    \theta_i
    -\Delta\theta\,
    \frac{\int_0^\lambda g_\epsilon(u)\,du}{\int_0^1 g_\epsilon(u)\,du},
    \qquad
    g_\epsilon(\lambda)=\exp\!\left[-\frac{\epsilon}{\lambda(1-\lambda)}\right],
\end{equation}
with \(\lambda=t/T\),
\(\Delta\theta=\theta_i-\theta_f\),
and \(\epsilon\) chosen so that \(|\dot\theta|<2h_x\).
The corresponding exact physical correction is
\begin{equation}
    c_z(t)
    =
    \frac{1}{2}\frac{d}{dt}\arcsin\!\left[-\frac{\dot\theta(t)}{2h_x}\right]
    +
    h_x\cot\theta(t)\sqrt{1-\frac{\dot\theta(t)^2}{4h_x^2}}
    - h_z(t).
\end{equation}

More broadly, perfect or near-perfect transfer in Landau-Zener-type settings can also be engineered by exploiting nonadiabatic interference, including Landau-Zener-St\"uckelberg interferometry and related composite-pulse strategies~\cite{Shevchenko2010Landau,Ivakhnenko2023Nonadiabatic,Torosov2011Smooth,Ryzhov2024Alternative}.
Those approaches cancel transitions at the level of the physical dynamics, whereas the present construction produces an exact endpoint-matched \(Z\)-only dressed-CD correction directly from the commutator equation.
For comparison,
approximate scalar sweep shaping in the Landau-Zener setting,
such as fast quasiadiabatic dynamics (FAQUAD)~\cite{Martinez-Garaot2015Fast},
improves adiabatic transport by redesigning the detuning profile but does not produce an exact CD correction.

\begin{figure}[t]
\centering
\includegraphics[width=\linewidth,draft=false]{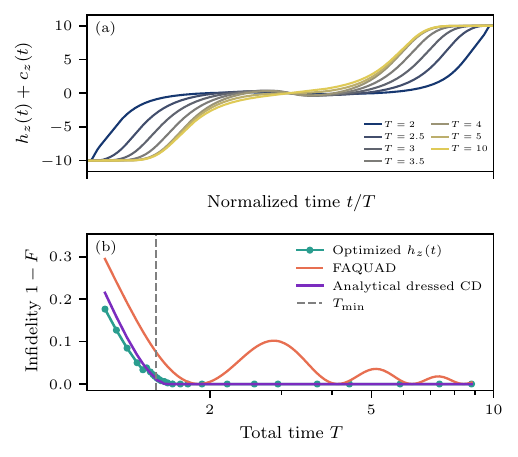}
\caption{
Exact \(Z\)-only dressed-frame solution for the Landau-Zener problem.
(a)
Implemented detuning profiles \(h_z(t)+c_z(t)\) from the exact dressed-frame construction for representative total durations \(T\).
The small central bump is part of the exact solution family:
it fine-tunes the detuning near the middle of the sweep so that the dynamics follows the dressed adiabatic trajectory while satisfying the endpoint constraints.
(b)
Direct optimization over detuning-only controls, shown as the final infidelity \(1-F\) versus total time \(T\), together with the analytical speed-limit threshold \(T_{\min}\).
The analytical solution is exact for \(T>T_{\min}\) and is capped below the threshold, where no regular \(Z\)-only solution exists.
Above \(T_{\min}\),
the optimization reaches floating-point exact transport.
}
\label{fig:variational-lz-z-only}
\end{figure}

\Cref{fig:variational-lz-z-only} illustrates both features of the construction.
The upper panel shows that the exact \(Z\)-only solution is not a trivial monotone reshaping of the bare detuning.
Especially near the minimum-time regime,
the implemented profile develops a small midpoint bump that provides the final adjustment needed for exact transport.
The lower panel shows the corresponding quantum speed limit.
The slight rise visible just before \(T_{\min}\) is due to the finite-sampling of the simulation near this sharp threshold,
not a breakdown of the exact construction above the bound.

\subsection{Reduction of the Bell-state example to the Landau-Zener problem}
\label{app:bell-lz-reduction}

Within the invariant subspace spanned by
\(\ket{\Phi^+}=(\ket{00}+\ket{11})/\sqrt{2}\)
and
\(\ket{\Psi^+}=(\ket{01}+\ket{10})/\sqrt{2}\),
the operators used in the Bell-state example act as
\begin{equation}
    \begin{aligned}
        Z_1 Z_2 &\mapsto Z, \\
        X_s &\mapsto \sqrt{2}X, \\
        ZY_s &\mapsto \sqrt{2}Y .
    \end{aligned}
\end{equation}
Introducing the relabeled effective Pauli operators
\begin{equation}
    \begin{aligned}
        X_{\mathrm{eff}}&=-Z, \\
        Y_{\mathrm{eff}}&=Y, \\
        Z_{\mathrm{eff}}&=X,
    \end{aligned}
\end{equation}
the reference Hamiltonian and constrained correction become
\begin{equation}
    \begin{aligned}
        H_0(t)&=|J|X_{\mathrm{eff}}+\sqrt{2}\,h_x(t)Z_{\mathrm{eff}}, \\
        H_c(t)&=\sqrt{2}\,c_x(t)Z_{\mathrm{eff}} .
    \end{aligned}
\end{equation}
This is exactly the same two-level problem as the Landau-Zener \(Z\)-only construction above, with the replacements
\begin{equation}
    \begin{aligned}
        h_x &\mapsto |J|, \\
        h_z(t) &\mapsto \sqrt{2}\,h_x(t), \\
        c_z(t) &\mapsto \sqrt{2}\,c_x(t).
    \end{aligned}
\end{equation}

Under this identification, the analytical family derived in \cref{app:lz-z-only} applies verbatim.
In particular, the reduced two-level problem obeys the bound
\begin{equation}
    T \ge \frac{|\theta_f-\theta_i|}{2|J|},
\end{equation}
where
\begin{equation}
    \cos\theta_i=
    \frac{\sqrt{2}\,h_x(0)}{\sqrt{J^2+2h_x(0)^2}},
    \qquad
    \cos\theta_f=
    \frac{\sqrt{2}\,h_x(T)}{\sqrt{J^2+2h_x(T)^2}}.
\end{equation}
For the Bell-state preparation problem itself, it also applies the physical requirement
\begin{equation}
    T \ge \frac{\pi}{4|J|},
\end{equation}
because the fixed \(Z_1Z_2\) coupling is the sole entangling resource and must act long enough to accumulate the required entangling phase.
Likewise, every admissible choice of \(\theta(t)\) yields an exact \(X_s\)-only correction through
\begin{equation}
    c_x(t)=\frac{1}{\sqrt{2}}\,c_z^{\mathrm{LZ}}(t),
\end{equation}
where \(c_z^{\mathrm{LZ}}(t)\) is the Landau-Zener expression after the replacements above.

For the analytical traces shown in \cref{fig:variational-ising-2q-x-only},
we use the endpoint-flat bump-density family inherited from the Landau-Zener construction.
This family is endpoint-flat to all orders and remains exact whenever
\begin{equation}
    \max_t |\dot\theta(t)| < 2|J|.
\end{equation}
In practice, for a fixed total time \(T\), we choose \(\epsilon\) so that the flatter profile stays safely within the admissible region,
\begin{equation}
    \max_t |\dot\theta(t)| \lesssim 0.98\times 2|J|,
\end{equation}
which is the condition used for the analytical traces in \cref{fig:variational-ising-2q-x-only}.
The pulse \(c_x(t)\) constructed in this way is then inserted back into the full two-qubit Hamiltonian
\begin{equation}
    H(t)=J Z_1Z_2+\bigl[h_x(t)+c_x(t)\bigr]X_s,
\end{equation}
and the population and infidelity curves in \cref{fig:variational-ising-2q-x-only} are obtained by evolving that full four-level model rather than the reduced two-level one.

\section{Tripod implementation details}
\label{app:tripod-details}

For the tripod example in \cref{fig:variational-tripod},
the instantaneous dark states of the bare loop are
\begin{equation}
    \ket{D_1(t)}
    =
    -\sin\phi(t)\ket{0}
    +
    \cos\phi(t)\ket{1},
\end{equation}
\begin{equation}
    \ket{D_2(t)}
    =
    \cos\theta(t)(
    \cos\phi(t)\ket{0}
    +
    \sin\phi(t)\ket{1})
    -
    \sin\theta(t)\ket{a}.
\end{equation}
The target logical gate is
\begin{equation}
    U_{\mathrm{tar}}
    =
    e^{-i\pi \sigma_y/4}
    =
    R_y(\pi/2)
\end{equation}
written in the ordered dark basis \(\{\ket{D_1},\ket{D_2}\}\).
If \(U(T)\) is the full \(4\times 4\) propagator and \(B_i\), \(B_f\) are the matrices whose columns are the initial and final dark states,
the logical block used in the benchmarks is
\begin{equation}
    U_{\mathrm{logical}}=B_f^\dagger U(T) B_i .
\end{equation}
We report the leakage
\begin{equation}
    L
    =
    1-\frac{\operatorname{Tr}(U_{\mathrm{logical}}^\dagger U_{\mathrm{logical}})}{2},
\end{equation}
the subspace fidelity
\begin{equation}
    F_Q
    =
    \frac{\left|\operatorname{Tr}(U_{\mathrm{logical}}U_{\mathrm{tar}}^\dagger)\right|^2}
    {2\operatorname{Tr}(U_{\mathrm{logical}}^\dagger U_{\mathrm{logical}})},
\end{equation}
and the total fidelity
\begin{equation}
    F
    =
    \frac{2}{3}F_Q(1-L)+\frac{1-L}{3}.
\end{equation}

The physical correction basis for the tripod solution is restricted to the native star-graph operators
\(\{X_{e0},X_{e1},X_{ea}\}\),
while the virtual dressed-frame generator is expanded in
\(\{Y_{01},Y_{0a},Y_{1a},X_{e0},X_{e1},X_{ea}\}\).
All coefficient functions are represented in finite Fourier Sine bases and optimized with L-BFGS-B against the commutator residual.
For the robustness panels,
we use \(\Omega=1\),
\(T=25\),
\(\theta(t)=\theta_{\max}\sin^3(\pi t/T)\),
\(\phi(t)=2\pi t/T\),
and \(\theta_{\max}\) is initialized at \(2.73\),
calibrated so that the bare holonomy matches \(R_y(\pi/2)\).
The robustness solution uses \(14\) Fourier-sine modes.
The adiabatic reference in \cref{fig:variational-tripod}(a,b) uses the same loop with \(T=200\) and no CD correction.

For \cref{fig:variational-tripod}(a),
the common amplitude perturbation rescales all three tripod legs as
\begin{equation}
    (\Omega_0,\Omega_1,\Omega_a)\to (1+\epsilon)(\Omega_0,\Omega_1,\Omega_a),
\end{equation}
with \(\epsilon\in[-0.05,0.05]\) sampled at \(11\) equally spaced points.
For \cref{fig:variational-tripod}(b),
the detuning perturbation adds \(\Delta_e\ket{e}\bra{e}\) with
\(\Delta_e/\Omega\in[-0.3,0.3]\) sampled at \(13\) equally spaced points.
The direct benchmark is an external two-level comparison on the \(\{\ket{0},\ket{1}\}\) manifold with
\begin{equation}
    H_{\mathrm{dir}}=\Omega Y_{01},
    \qquad
    Y_{01}=i(\ket{1}\bra{0}-\ket{0}\bra{1}),
\end{equation}
chosen so that the ideal pulse implements \(R_y(\pi/2)\).
For the amplitude test,
the same benchmark uses \(\Omega\to (1+\epsilon)\Omega\).
For the detuning test,
we use
\begin{equation}
    H_{\mathrm{dir},\Delta}
    =
    \Omega Y_{01}
    +
    \frac{\Delta}{2} Z_{01},
    \qquad
    Z_{01}=\ket{0}\bra{0}-\ket{1}\bra{1},
\end{equation}
with fixed on-resonance pulse duration \(\pi/(4\Omega)\).

For \cref{fig:variational-tripod}(c),
the bare and dressed protocols are compared over \(25\) logarithmically spaced gate times \(T\in[10,200]\).
At each \(T\),
the \(\theta_{\max}\) is recalibrated.
The time-sweep solve uses \(20\) Fourier-sine modes.

\end{document}